\documentclass[12pt]{iopart}

\usepackage{graphicx}
\usepackage{epsfig,epstopdf}
\usepackage{caption}
\usepackage{subfigure}
\begin{document}
\textbf{
\title{Suppression of a laminar kinematic dynamo by a prescribed large-scale shear}
}

\author{Aditi Sood${ˆ1}$, Rainer Hollerbach{$ˆ2$}, Eun-jin Kim{$ˆ1$}}
\address{$ˆ1$ School of Mathematics \& Statistics, University of Sheffield,
Sheffield S3 7RH, UK}
\address{$ˆ2$ School of Mathematics, University of Leeds, Leeds LS2 9JT, UK}

\begin{abstract}
We numerically solve the magnetic induction equation in a spherical shell
geometry, with a kinematically prescribed axisymmetric flow that consists of
a superposition of a small-scale helical flow and a large-scale shear flow.
The small-scale flow is chosen to be a local analog of the classical Roberts
cells, consisting of strongly helical vortex rolls. The large-scale flow is a
shearing motion in either the radial or the latitudinal directions. In the
absence of large-scale shear, the small-scale flow operates
{ very effectively}
as a dynamo, in agreement with previous results. Adding increasingly large
shear flows strongly suppresses the dynamo efficiency, indicating that shear
is not always a favourable ingredient in dynamo action.
\end{abstract}

\maketitle

\section{Introduction}
Many astrophysical objects such as planets or stars possess magnetic fields.
The origin of all of these fields is believed to be via so-called dynamo
action, whereby the motion of electrically conducting fluid maintains the
field against the otherwise inevitable ohmic decay. The full problem involves
nonlinearly coupled partial differential equations for the evolution of both
the magnetic field and the fluid flow, with each affecting the other. Despite
the complexity of this process, numerical solutions have by now progressed to
the point where quite detailed and realistic models of different dynamos are
routinely being computed \cite{Jones,Brandenburg,Ferrario}.
{ Although dynamos in different astrophysical objects are often
very different in many important details, there are nonetheless a few basic
ingredients that occur often enough to warrant studying their effects in
isolation.} Two of these are small-scale helical
motions and large-scale shear flows. In this work we will study a
kinematically prescribed flow that consists of a superposition of these two.
Although simply prescribing the flow, rather than self-consistently solving
for it, is of course a simplification of the full problem, this approach has
a considerable merit of elucidating the key role of helicity and shear flows
and the interaction between different competing effects, which are often
difficult to differentiate in a more complete model.

The idea that the helicity, or handedness, of a flow can be a critically
important ingredient in dynamo theory is familiar since the so-called
mean-field theory was developed in the 1960s (e.g.\ \cite{Moffatt}), in
which certain averages are taken, and an $\alpha$-effect is extracted that
is directly proportional to the helicity of the small-scale flow structures.
This $\alpha$-effect can then drive a large-scale dynamo. One particularly
simple flow for which the theory can be thoroughly developed is the Roberts
flow \cite{roberts1970,roberts1972}, consisting of a plane-periodic array of
helical vortices. A time-periodic version of the Roberts flow \cite{Galloway}
has also proven invaluable in studying the distinction between slow and fast
dynamos; a necessary condition for fast dynamo action, in which the magnetic
field grows on the fast advective timescale rather than the slow diffusive
timescale (or any intermediate timescale) is that Lagrangian particle paths
be chaotic \cite{Childress}, which in a flow that depends on $x$ and $y$ but
not $z$ requires the flow to also depend on $t$. Another adaptation of a
Roberts-type flow was by \cite{richardson2012}, who fitted a similar array of
small-scale cells into a spherical shell. The motivation for this is to
introduce a natural largest scale, unlike in plane-periodic geometry, where
the largest scale is effectively infinite.

Another ingredient that can play an important role both in dynamo theory
generally as well as in particular situations such as the solar tachocline
\cite{tachoclineref}, accretion disks \cite{accdiskref} or entire galaxies
\cite{galaxyref} is a large-scale shear. At its most basic, in what is known
as the $\omega$-effect, a shear flow can simply take an existing magnetic
field and stretch it out, thereby producing a field component along the
direction of the shear. A vast number of numerical models
{
\cite{rogachevskii2003,rogachevskii2004,brandenburg2008,kapyla2008,
yousef2008a,yousef2008b,kapyla2009,hughes2009,hughes2013,heit2011,mitra2012}
include various large-scale shearing motions, and find that it enhances the
dynamo efficiency, in flows both with and without helicity.}
However, numerous
other studies \cite{leprovost2008,leprovost2009,KIM3,Cour}
find that shear can also be detrimental to dynamo action, for example by
disrupting critically important phase relationships between different
small-scale cells. Yet another study \cite{tobias2013} finds that shear may
enhance large-scale fields but suppress small-scale ones.

It is these conflicting possibilities regarding the effect of a large-scale
shear that motivate our work. Specifically, we start with a spherical
shell cellular flow similar to that of \cite{richardson2012}, add to it
large-scale shear flows of increasing strength, and consider the dynamo
action of these kinematically prescribed flows. By examining the growth rate
curves, as functions of magnetic Reynolds number, we show that at least for
these small-scale flows, the addition of a large-scale shear always suppresses
the dynamo efficiency. We also examine the spatial structures of the
resulting eigenmodes, as well as the corresponding magnetic energy spectra,
and explore the influence of the shear on these.

\section{Governing Equations}
In the framework of kinematic dynamo theory, we solve the induction equation
\begin{equation}
\frac{\partial \mathbf{B}}{\partial t} =
 \nabla\times(\mathbf{U}\times \mathbf{B})
 + R_m^{-1}\nabla^{2}\mathbf{B}
\label{induction}
\end{equation}
in a spherical shell. The magnetic Reynolds number $R_m=UL/\eta$, where $U$ is
a characteristic velocity scale, $L$ a characteristic length scale (taken to be
the outer radius of the shell), and $\eta$ is the magnetic diffusivity of the
fluid. The prescribed velocity field $\mathbf{U}$ is axisymmetric, and consists
of a super-position of a small-scale helical flow and a large-scale shear flow.

The small-scale flow has the form
$\nabla\times(\psi\mathbf{\hat{e}_{\phi}})
              +  v\mathbf{\hat{e}_{\phi}}$,
where the meridional circulation $\psi$ and
the azimuthal velocity $v$ are given by
\begin{eqnarray}
\psi=\frac{1}{N_{\theta}}r
 \sin\left(\frac{(r-r_{i})}{(r_{0}-r_{i})}N_{r}\pi\right)
 \sin\theta \cos\theta \cos(N_{\theta}\theta),\\
v  = \sin\left(\frac{(r-r_{i})}{(r_{0}-r_{i})}N_{r}\pi\right)
 \sin\theta \cos(N_{\theta}\theta).
\end{eqnarray}
This flow is very similar to one of the flows considered by
\cite{richardson2012}, and consists of small-scale cells that are local analogs
of the classical Roberts flow \cite{roberts1970,roberts1972}. The parameters
$N_r$ and $N_\theta$ specify the number of cells in the $r$ and $\theta$
directions, respectively; we will here always take $N_\theta=4N_r$, which yields
cells that are very close to round. Note also that these flows have $O(1)$
magnitude, and hence a turnover time $\sim N_r^{-1}$ for the small cells.

The main difference between this flow and the previously considered
\cite{richardson2012} flow is the additional factor of $\cos\theta$ in $\psi$.
Without this factor the helicity of the flow would be equatorially symmetric;
with it included we see that $\psi$ is anti-symmetric, $v$ is symmetric, and
the helicity is then also anti-symmetric. There are two reasons for modifying
the flow in this way. First, it is simply of interest to see whether the
previous results \cite{richardson2012} continue to hold even if the small-scale
flow is only strongly helical in each hemisphere separately, but with zero net
helicity.  Second, taking $\psi$ to be anti-symmetric and $v$ symmetric allows
for the familiar separation of $\mathbf{B}$ into dipole and quadrupole
symmetries, and is thus numerically convenient.

Turning next to the large-scale shear flow, this is of the form
$S\,r\sin\theta\,\Omega$, where the two choices for the angular velocity
$\Omega$ are
\begin{equation}
\Omega_1 = (r - 0.75),\qquad\qquad \Omega_2=(\cos^2\theta - 0.5).
\end{equation}
That is, $\Omega_1$ represents a shear purely in the radial direction, whereas
$\Omega_2$ represents a shear purely in the latitudinal direction (but still
equatorially symmetric, to preserve the dipole/quadrupole decoupling for
$\mathbf{B}$). The constants in each case, 0.75 and 0.5, correspond to
solid-body rotation, and hence have no effect other than to choose a
coordinate system where the average value of $\Omega$ is comparatively small.
{
We emphasize also that both the small-scale and the large-scale flows are
simply prescribed, rather than being solutions of the Navier-Stokes equation.
If $\mathbf{U}$ were evolved according to the Navier-Stokes equation, both
components would inevitably be far more complicated, and furthermore each
would significantly affect the other. By kinematically prescribing $\mathbf{U}$
we are able to avoid this mutual interdependence, and thereby isolate the
effect of increasing shear on the dynamo action of the small-scale flow.
}

To summarize, we solve Eq.~(\ref{induction}) in a spherical shell with radii
$r_i=0.5$ and $r_o=1$, with the total flow given by
\begin{equation}
\mathbf{U}=\nabla\times(\psi\mathbf{\hat{e}_{\phi}})
         +  v\mathbf{\hat{e}_{\phi}} + S\,r\sin\theta\,\Omega,
\end{equation}
and $\psi$, $v$ and $\Omega$ as above. Each choice of flow, {Flow1 in the presence of radial shear}, $\Omega_1$, and
{Flow2 in the presence of latitudinal shear}, $\Omega_2$, is therefore completely specified by the two parameters $N_r$,
measuring the number of cells in the radial direction, and $S$, measuring the
strength of the large-scale shear.  (Note incidentally that according to this
definition the magnetic Reynolds number $R_m$ is based on the magnitude of the
small-scale flow; a `large-scale flow Reynolds number' would require
multiplying $R_m$ by $S$.)

Because the flow $\mathbf{U}$ is axisymmetric, $\mathbf{B}$ decouples into
distinct $\exp(im\phi)$ azimuthal modes, each of which further decouple into
dipole and quadrupole equatorial symmetries as noted above. In fact, the two
symmetries always behaved very similarly, so only dipole results will be
presented in detail here. The further details of the numerical solution are
exactly as in \cite{richardson2012}, see also \cite{Rainer2000}. Resolutions
up to 300 Chebyshev polynomials in $r$ and 400 Legendre functions in $\theta$
were used, and were carefully checked to ensure fully resolved solutions.
{
See Table 1 for sample convergence results at different resolutions.
}
Typical time-steps used were $O(10^{-3})$, and all solutions were run
sufficiently long to allow the dominant eigenmode to emerge.

{
\begin{table}[h]
\begin{tabular}{c|c|c|c|c|c|c|c}
$\;N_r\;$&$\;S_1\;$&$\;S_2\;$&$120\times240$&$140\times280$&
 $160\times320$&$180\times360$&$200\times400$\\ \hline
 5 & 7 & 0 & 0.848 & 0.824 & 0.787 & 0.770 & 0.762 \\
 5 & 0 & 7 & 0.681 & 0.604 & 0.563 & 0.543 & 0.530 \\
10 & 7 & 0 & 1.054 & 1.047 & 1.042 & 1.038 & 1.036 \\
10 & 0 & 7 & 1.293 & 1.201 & 1.267 & 1.228 & 1.245 \\
  \end{tabular}
\caption{Growth rates for the $N_r=5$ and $10$ flows, with shear parameter $S=7$
for either $\Omega_1$ or $\Omega_2$, as indicated. For the five resolutions given,
the first quantity is the number of Chebyshev polynomials in $r$, and the second
is the number of Legendre functions in $\theta$. The azimuthal wavenumber $m=10$
for all four rows; other values of $m$ have similar convergence properties.
$R_m=10^4$ for all results; smaller values typically converged even more quickly.}
\label{table1}
\end{table}
}

\section{Results}

Figure 1 shows growth rates for the two cell sizes $N_r=5$ and 10, and
shear parameter $S=0$ (so small-scale cells only) and $S=3$ for the two
shearing options. The $S=0$ results are exactly as one might expect: for
increasing $R_m$ the growth rates first increase, then eventually decrease
again, as they must for a slow dynamo (since the flow is independent of time).
There is also a smooth progression to higher azimuthal wavenumbers $m$ being
the most dominant modes. These results are broadly similar to the previous
results \cite{richardson2012}, including the feature that $N_r=10$ has a
maximum growth rate significantly greater than $N_r=5$ does. This is because
the turnover time of a small cell decreases with $N_r$ while the effective
growth rate should be measured in units of turnover time, rendering the
effective growth rate (measured in turnover time units) comparable in
$N_r=5$ and 10 cases. Another conclusion is that flows such as these, with
strong helicity in each hemisphere, but zero net helicity, still behave much
the same as the previous flows with helicity of the same sign everywhere.
\begin{figure}[ht!]
\centering
\subfigure[No shear, $N_r =5$]{%
\includegraphics[width=0.33\textwidth]{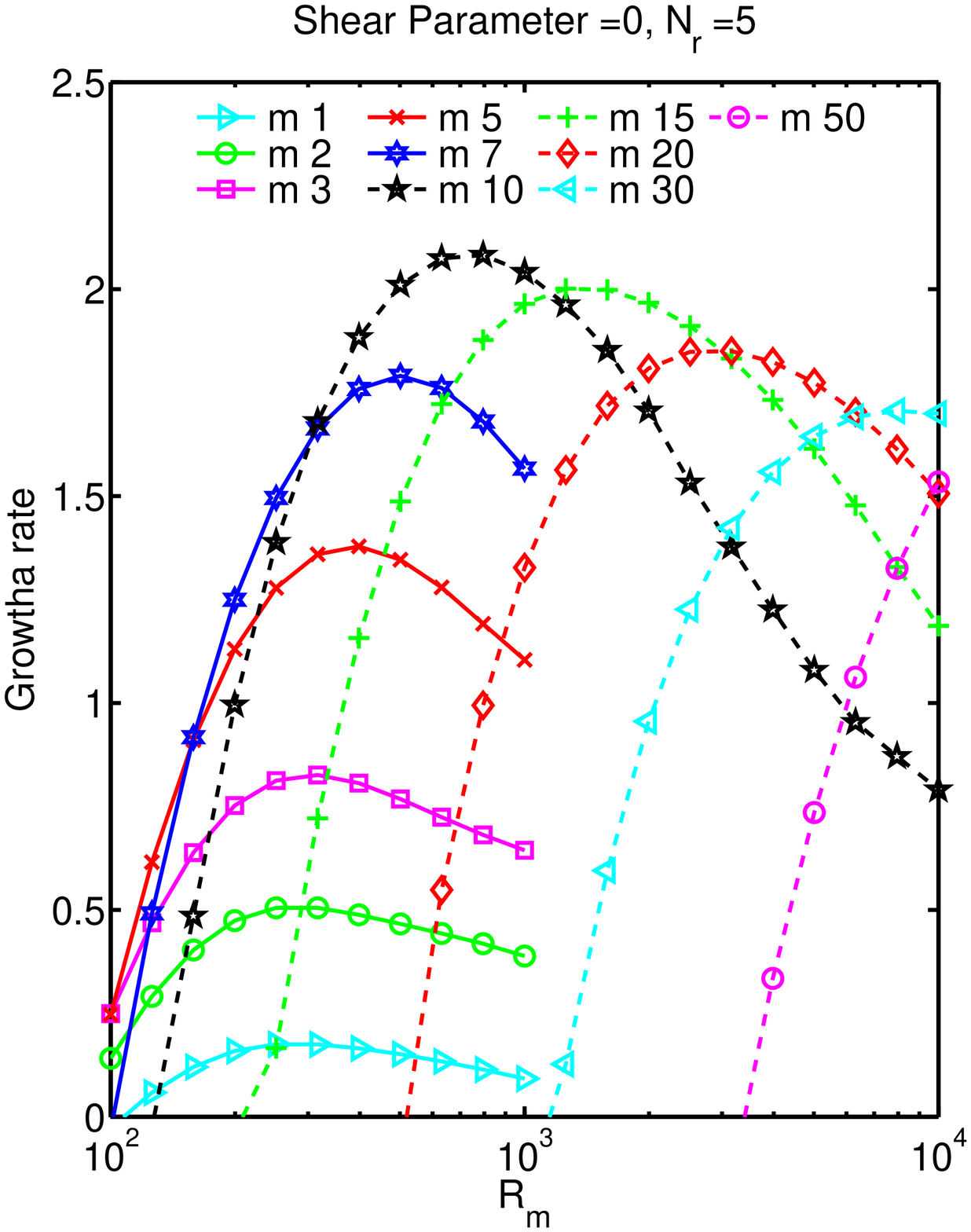}
}%
\subfigure[$S = 3$, \textbf{$\Omega_1$}, $N_r = 5$]{%
\includegraphics[width=0.33\textwidth]{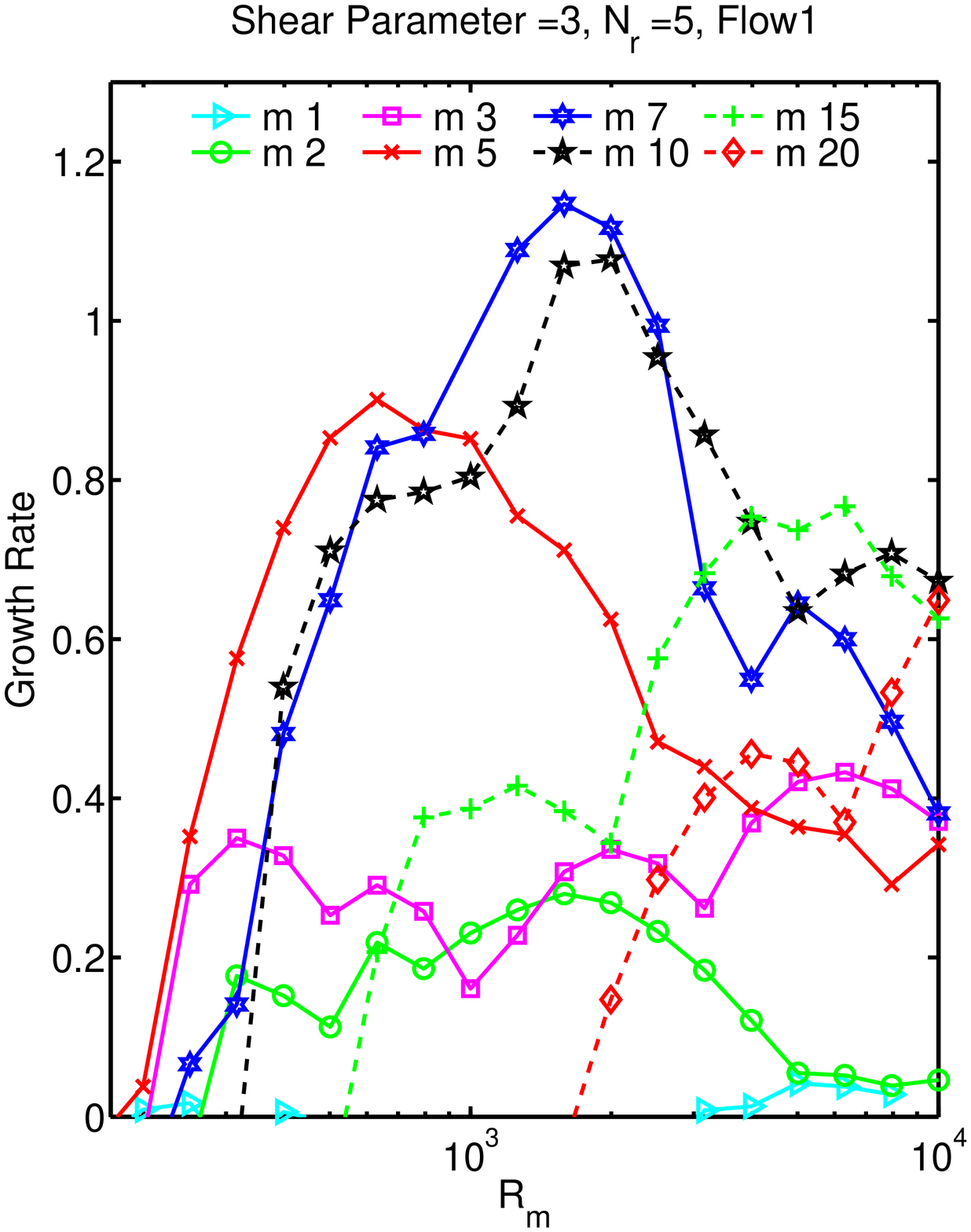}
}%
\subfigure[$S = 3$, \textbf{$\Omega_2$}, $N_r = 5$]{%
\includegraphics[width=0.33\textwidth]{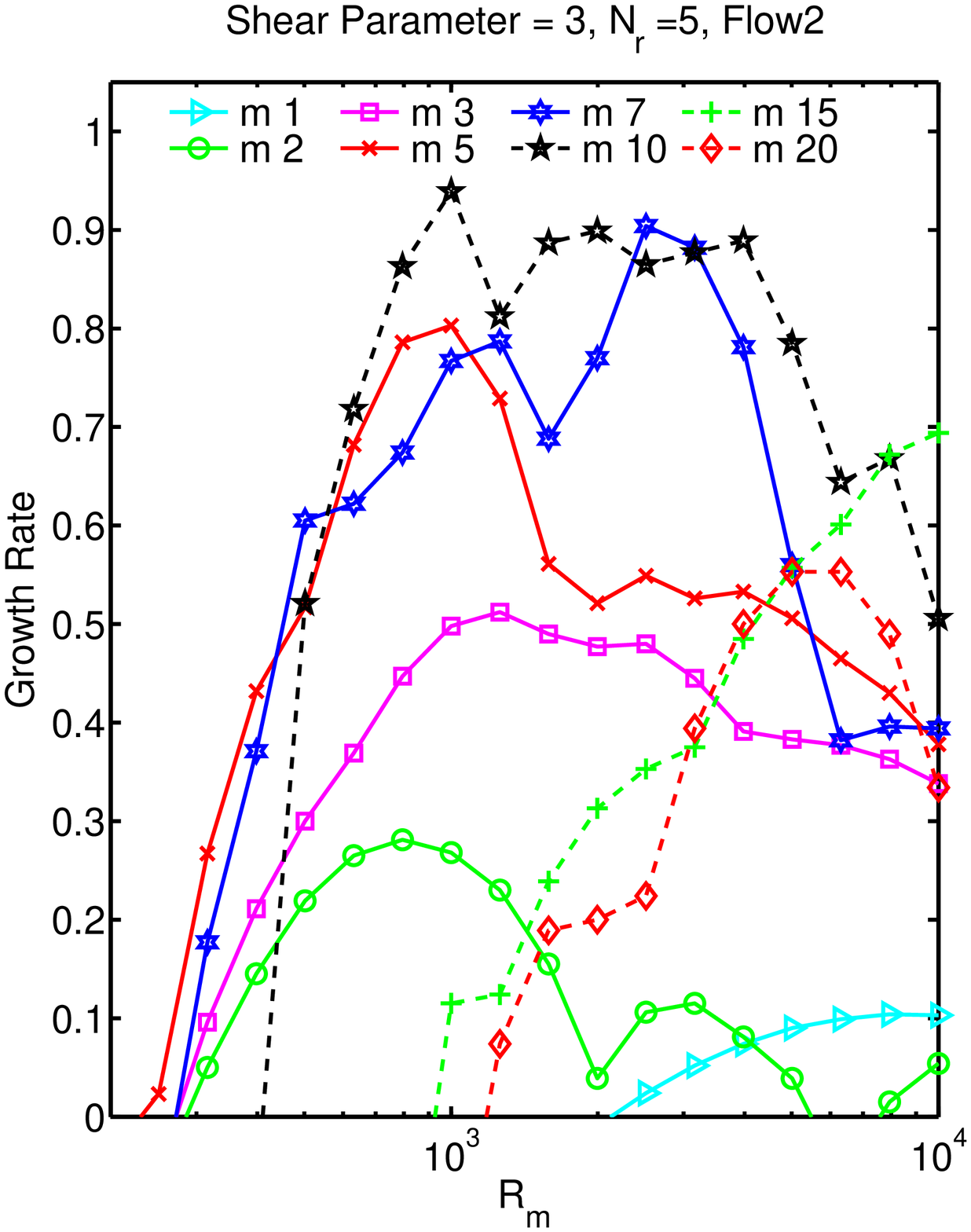}
}\\%
\subfigure[No shear, $N_r =10$]{%
\includegraphics[width=0.33\textwidth]{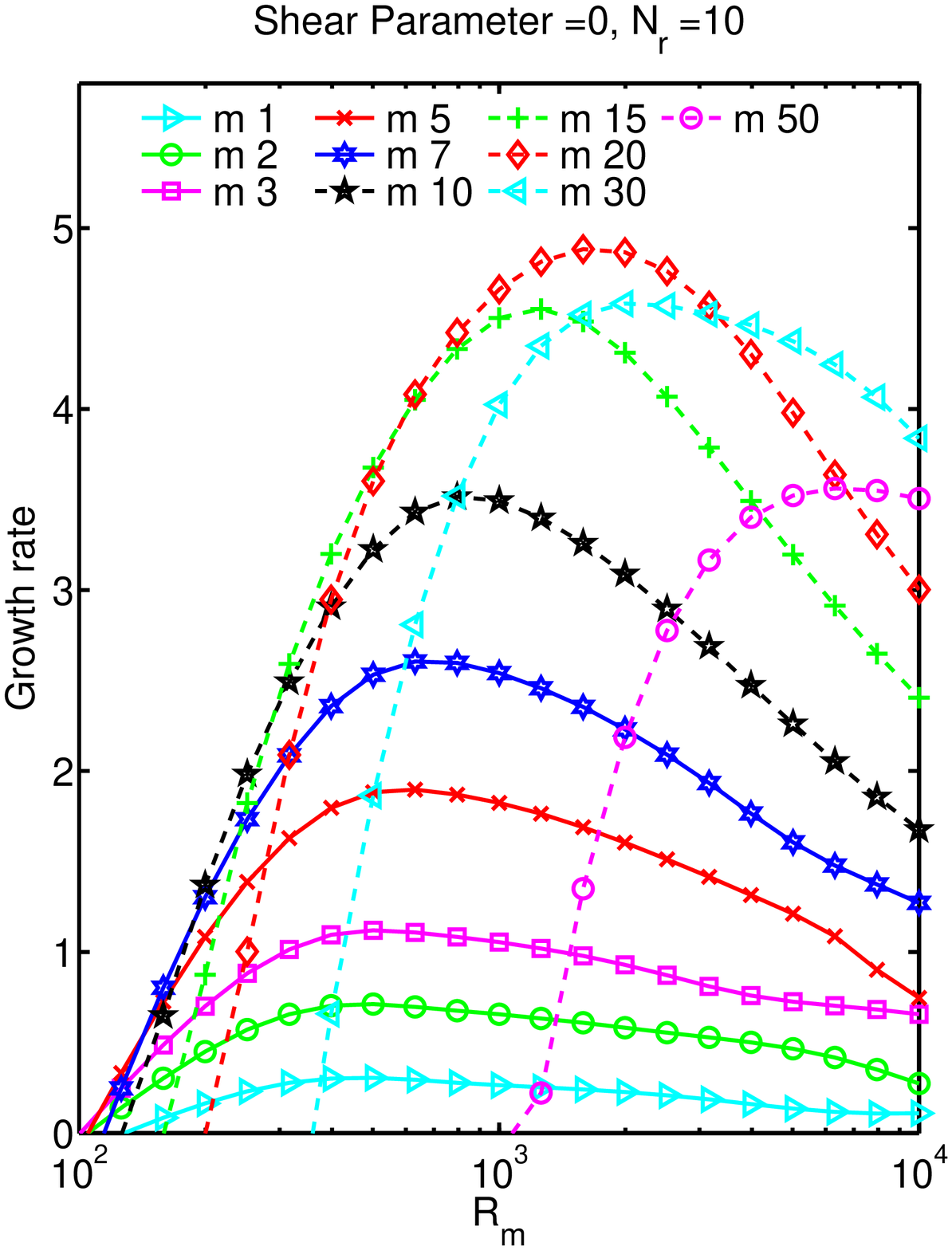}
}%
\subfigure[$S = 3$, \textbf{$\Omega_1$}, $N_r = 10$]{%
\includegraphics[width=0.34\textwidth]{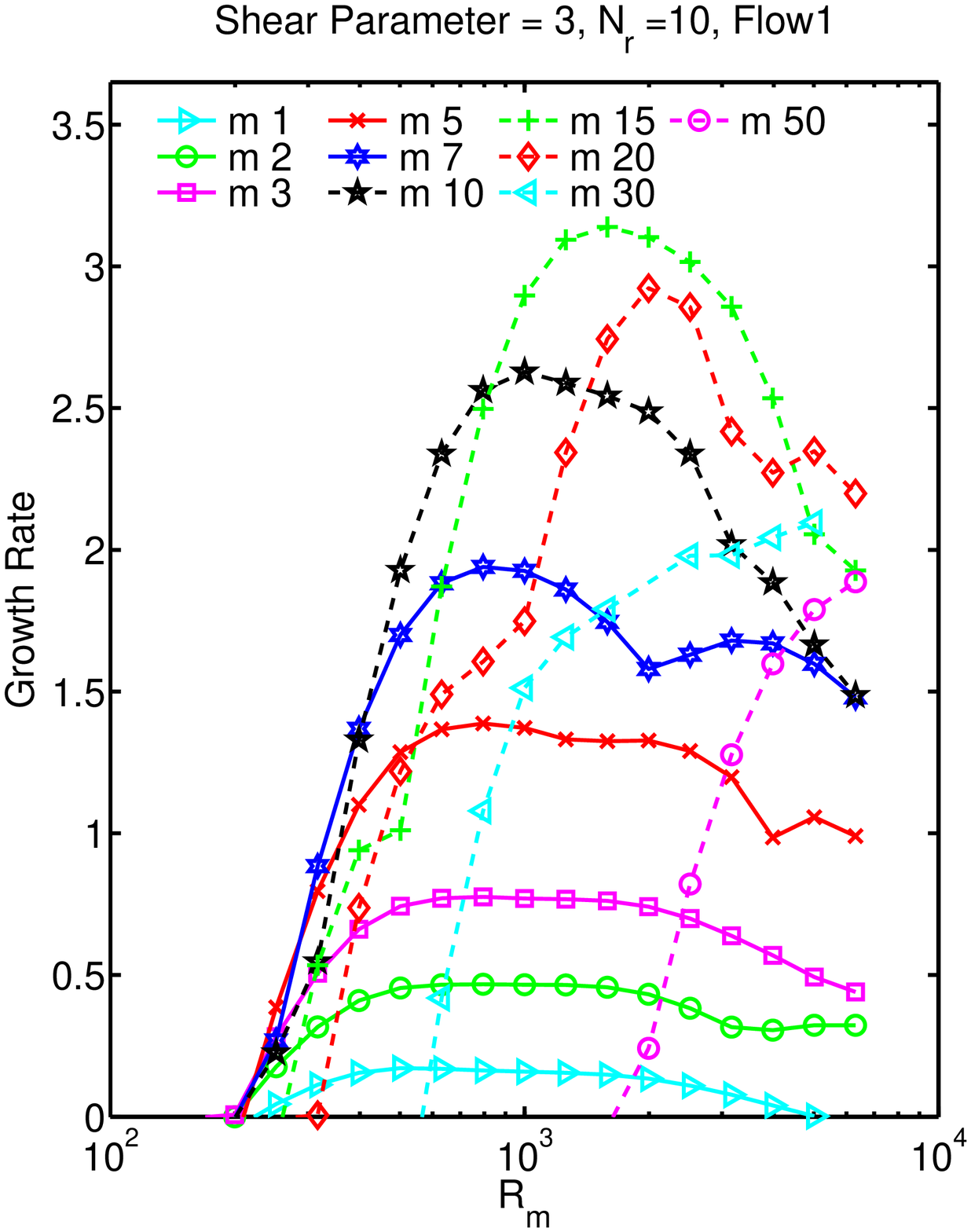}
}%
\subfigure[$S = 3$, \textbf{$\Omega_2$}, $N_r = 10$]{%
\includegraphics[width=0.34\textwidth]{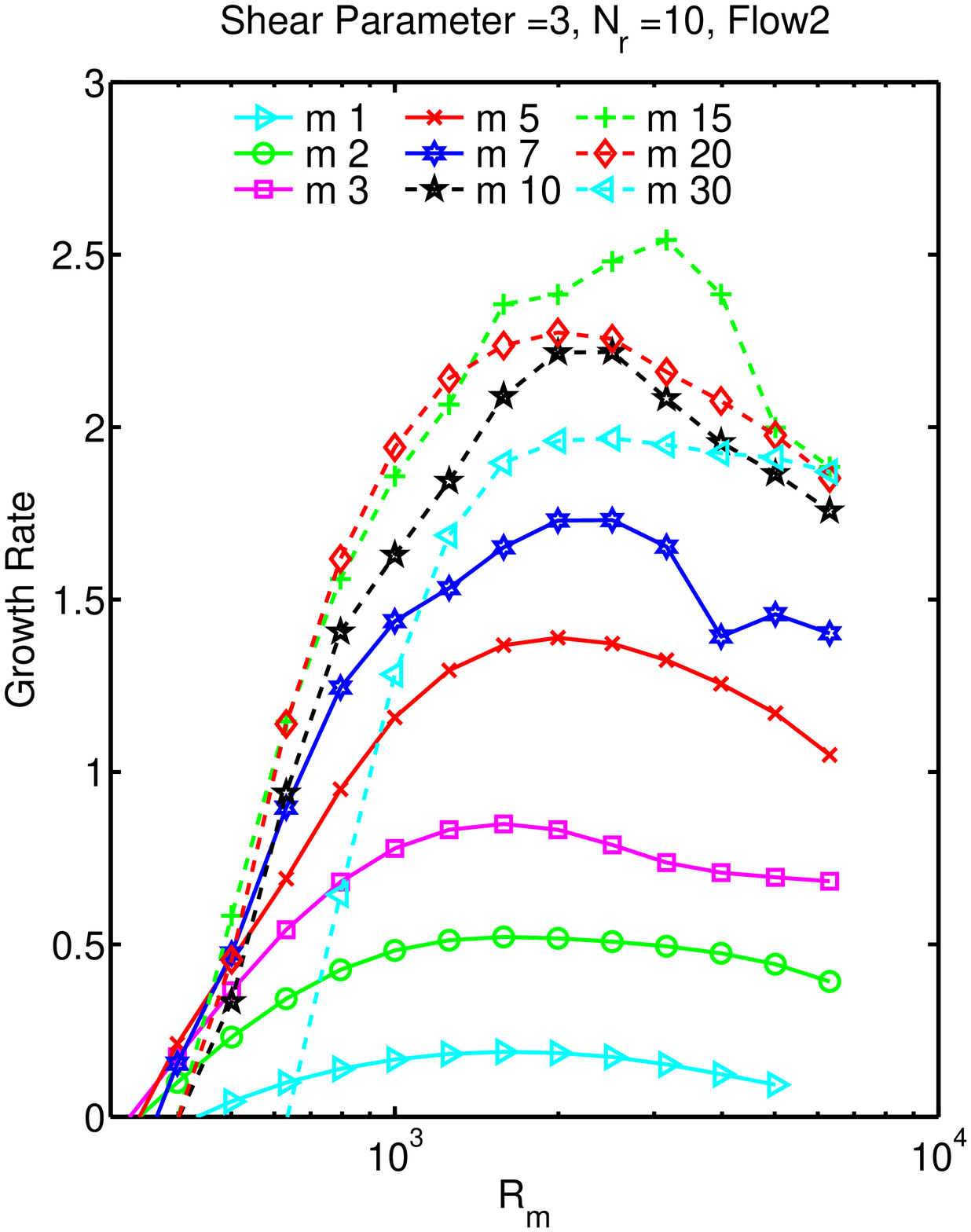}
}%
\caption{Growth rate curves as functions of $R_m$. The top row has $N_r=5$,
the bottom row $N_r=10$.
{
In each row the first panel has $S=0$, so no
large-scale shear at all, the second panel has the shear $\Omega_1$, and
the third panel has the shear $\Omega_2$, each with amplitude $S=3$.}}
\end{figure}
Turning next to the $S=3$ sheared cases, the most significant conclusion is
that the growth rates are strongly reduced in every case. Furthermore, $N_r=5$
curves are  far less smooth than they were for $S=0$. This seems to indicate
transitions between different eigenmodes due to shear flow -- for $S=0$ the
dominant eigenmode always remains the same mode, and just gradually evolves as
$R_m$ is increased. The $N_r=10$ curves also exhibit this mode switching to a
certain degree, but considerably less.

Specifically, in addition to the overall reduction in the growth rate, we see
a non-monotonic dependence of the growth rate on $R_m$, accompanied by the
shift of the location for the maximum growth rate and/or by the appearance of
peaks for secondary maxima. This irregular behaviour is more pronounced
in the case of  $N_r=5$ than $N_r=10$. In order to understand this, it is
useful to recall that a large-scale shear flow accelerates the formation of
small scales by distorting small scale structures, thereby facilitating the
dissipation of small scales by ohmic diffusion \cite{KIM1,leprovost2009}.

Quantitatively, the effective dissipation rate of small scales is given by
the decorrelation rate $1/\tau_\Delta$, by weighting the ohmic dissipation
rate $\gamma_\eta= k^2/R_m$ for the mode with wavenumber $k$ by shear
strength $S_*$ as follows \cite{KIM1}.
{ That is,
in the absence of a shear flow, the dissipation rate of a mode with wavenumber $k$
is given by $\gamma_\eta=k^2/R_m$. In the presence of a shear flow with an effective
shear parameter $S_*$, the dissipation rate of a $k$ mode is faster than $\gamma_\eta$
for sufficiently large $S_*$ as a shear flow accelerates the formation of small scales
which are then dissipated by Ohmic diffusion. This faster dissipation rate is estimated
by the decorrelation rate $\tau_\Delta^{-1}$ given by}
\begin{equation}
\tau_{\Delta}^{-1} = [\gamma_\eta\, S_*^2]^{1/3}
 = [\frac{k^2}{R_m} S_*^2]^{1/3}.
\label{shear1}
\end{equation}
In Eq. (\ref{shear1}), $S_*$ is the effective shear measured in the unit of
the characteristic time of small-scale cells
$$S_{*} = S/N_{r},$$
and $k$ is the wavenumber $k \sim m/0.75$ (by using the mean radius of the
shells). In the limit of strong shear $S_* > k^2/R_m$, the shear-weighted
decorrelation rate is much larger than the ohmic dissipation rate, leading to
the effectively {\it smaller} Reynolds number $R_m^{*}$, defined by the
following equality
\begin{equation}
\frac{k^2}{ R_m^{*}} =  \left[\frac{k^2}{R_m} \frac{S^2}{N_r^2}\right]^{1/3},
\label{shear2}
\end{equation}
or, alternatively
\begin{equation}
R_m = \left(\frac{R_m^{*}}{k}\right)^3 \frac{S^2}{k N_r^2}.
\label{shear3}
\end{equation}
Therefore, the $R_m$, which gives the same amount of the ohmic dissipation
for $R_m$ when $S=0$, increases with the shear strength $R_m>R_m^{*}$.
Although the flow and magnetic structures are far more complicated in our
model (e.g. compared to the Cartesian model \cite{Cour}), it is useful to
examine the consequence of Eq. (\ref{shear3}) by an order of estimate.
For instance, for $m=7$ and $N_r=5$ for which the maximum growth rate
appears around $R_m^{*} \sim 500$ without shear $S=0$ in Fig. 1(a), we can
estimate the value of $R_m$ when $S=3$ from Eq. (\ref{shear3}) by taking
$k \sim 10$ as
\begin{equation}
R_m \sim 4500,
\label{shear4}
\end{equation}
offering a possible explanation for the secondary local maximum growth
around $R_m \sim 4500$ in Fig. 1(b).

On the other hand, the dominant peak around $R_m\sim 2000$ seems to occur
since the growth rate measured in units of turnover time becomes comparable
to the decorrelation rate as
\begin{equation}
\frac{\gamma}{N_r} \sim \left[\frac{k^2}{R_m} \frac{S^2}{N_r^2}\right]^{1/3},
\label{shear4a}
\end{equation}
Solving Eq. (\ref{shear4a}) gives $R_m$ as
\begin{equation}
R_m \sim \frac{k^2 N_r S^2}{\gamma^3}.
\label{shear4b}
\end{equation}
For instance, for the $m=7$ ($k \sim 10$) mode discussed above,
taking $\gamma \sim 1.2$ and $S=3$, Eq. (\ref{shear4b}) yields $R_m\sim 2000$.

{
A key criterion for the quenching of dynamo by a shear flow is whether the effective
shear parameter $S_*$ is larger or smaller than the Ohmic dissipation rate
$\gamma_\eta = k^2/Rm$ for a mode with wavenumber $k$. Thus, other competing effects
are likely to promote a dynamo for small $S_*< \gamma_\eta$ while in the opposite
limit $S_* > \gamma_\eta$, a shear flow could inhibit a dynamo, dominating over
other effects.
}

Compared to $N_r=5$ case, the effective shear $S_*=S/N_r$ is smaller for
$N_r=10$, with much less effect of shear, and thus much smoother behaviour
in the growth rates in Figs. 1(d)-(e). Therefore, for even greater $N_r$, and
hence greater degree of separation between the small-scale and large-scale
flows, that the effect of shear is further reduced,  curves eventually
looking just as smooth as in the $S=0$ case. Further increasing $N_r$ would
unfortunately require prohibitively large numerical resolutions.

\begin{figure}[ht!]
\centering
\subfigure[]{%
\includegraphics[width=0.5\textwidth]{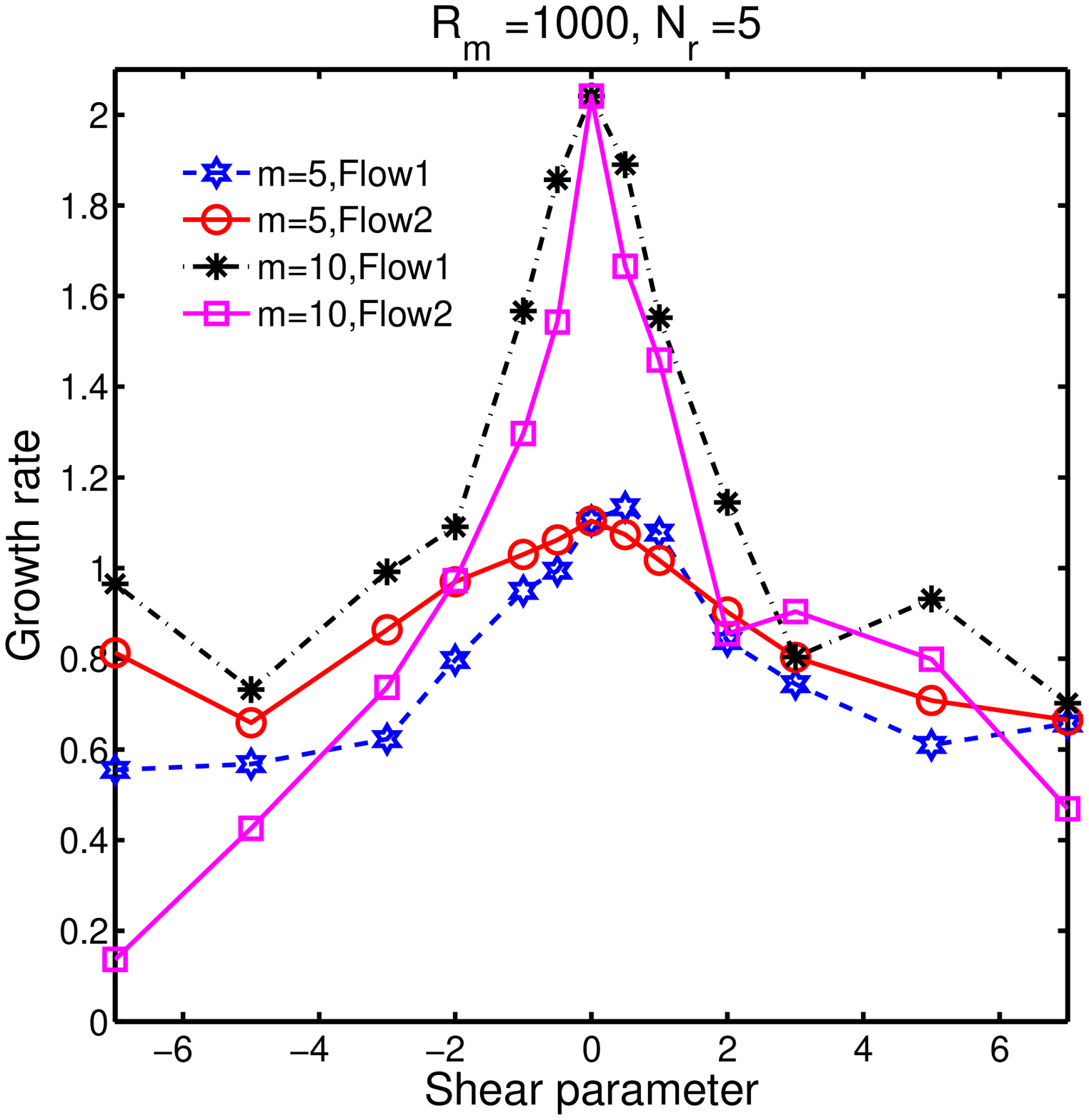}
}%
\subfigure[]{%
\includegraphics[width=0.5\textwidth]{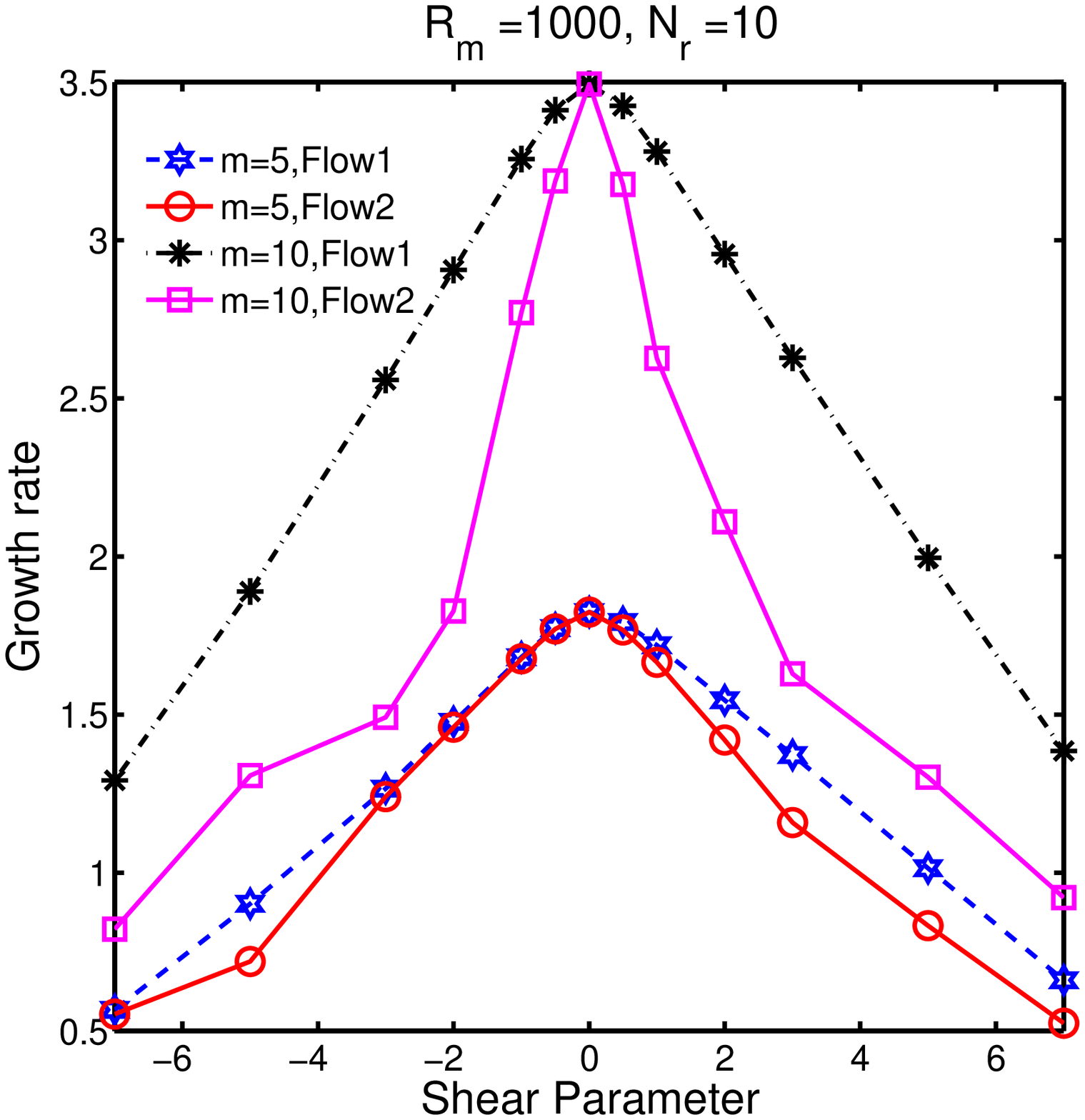}
}%
\caption{Growth rates as functions of shear parameter $S$, indicating the
suppression of dynamo action for non-zero $S$. The left and right panels have
$N_r=5$ and 10, respectively. \textbf{Flow1 in the presence of purely radial shear $\Omega_1$,  Flow2 in the presence of purely latitudinal shear $\Omega_2$}, and azimuthal wavenumbers $m=5$
and 10 as indicated, and $R_m=1000$ throughout.}
\end{figure}
Figure 2 quantifies the suppression of the growth rates by increasingly large
shear in the two cases. Detailed results are presented here only for the
particular value $R_m=1000$, close to the peaks in Fig. 1. Other values of
$R_m$ behaved qualitatively the same. Similarly, only the two azimuthal
wavenumbers $m=5$ and 10 are presented, but others were also examined and
behaved similarly. The overall conclusion is again very clear; even moderate
values of $S$ immediately begin to suppress the growth rates, almost in a
power law. As observed in Fig. 1, Fig. 2 also exhibits a narrow region of
$S$ with local maximum growth rate where the magnetic field is favourable to
the growth despite the overall growth inhibition. This seems to be caused by
resonance when the characteristic time scale ($1/\omega= 1/N_r$) of the
small-scale cells matches the local advection time for mode $k$ due to the
shear flow across the cell \cite{Cour} as
\begin{equation}
N_r \sim k  U_* = S (0.5/N_r),
\label{shear5}
\end{equation}
where the local mean flow across one single small cell is estimated as
$U_*=S (0.5/N_r)$ as there are $N_r$ cells between $r=[0.5,1]$. Note that
Eq. (\ref{shear5}) holds when the Doppler-shift frequency vanishes
($\omega - k U_*=0$). For instance, solving Eq. (\ref{shear5}) for $S$ for
the case $N_r=5$ and $m=10$ ($k\sim 15$), we obtain $S = 3 \sim 5$, explaining
the peak in Fig. 2(a). As is clear from Eq. (\ref{shear5}), the shear strength
$S$ required for the resonance increases quadratically with $N_r$, occurring
for much larger $S>10$ for $N_r=10$. This is why Fig. 2(b) shows only the
monotonic decrease in the growth rate with increasing $|S|$.

It is interesting
also to note that positive and negative values of $S$ yield similar results,
but not identical. This indicates that even for $N_r=10$ the cells are not so
small yet that the dynamo action of the overall pattern does not sense a
distinction between the inner and outer edges of the shell (for Flow 1), or
between more polar and more equatorial latitudes (for Flow 2)\\
\begin{figure}[ht!]
\centering
\subfigure[$N_r=5$, $S=0$]{%
\includegraphics[width=0.33\textwidth]{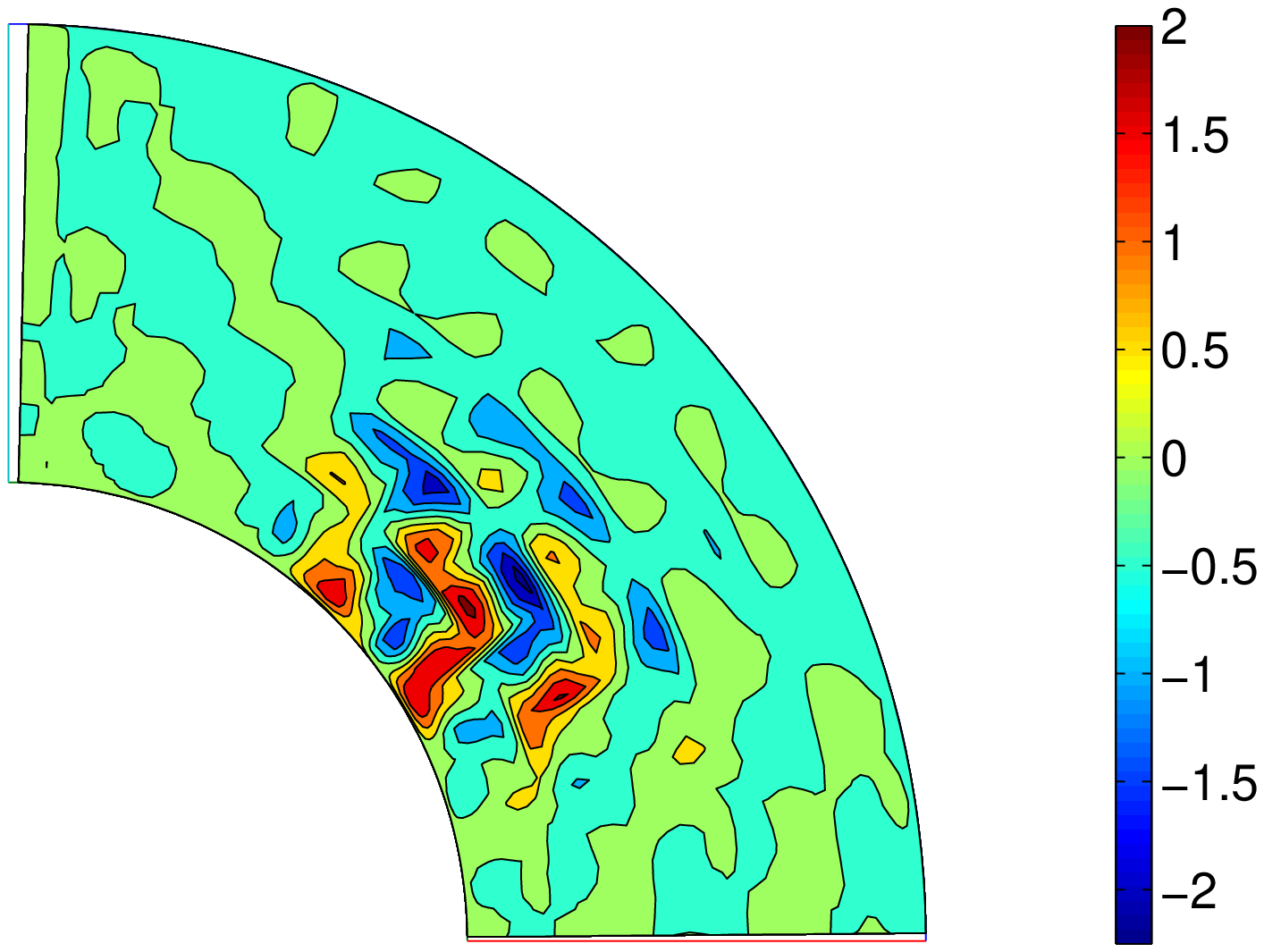}
}%
\subfigure[$N_r=5$, $S=3$, \textbf{$\Omega_1$}]{%
\includegraphics[width=0.33\textwidth]{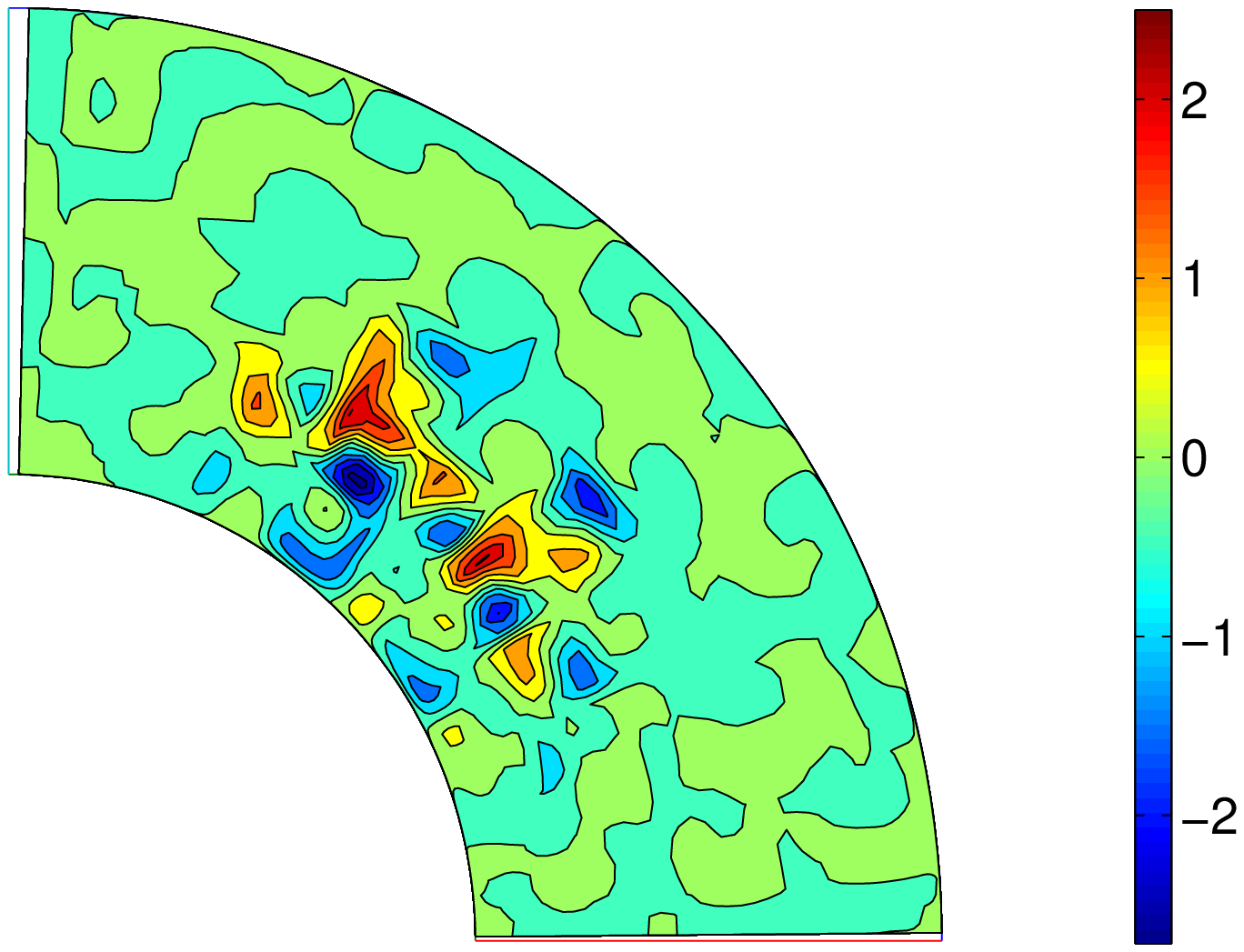}
}%
\subfigure[$N_r=5$, $S=3$, \textbf{$\Omega_2$}]{%
\includegraphics[width=0.33\textwidth]{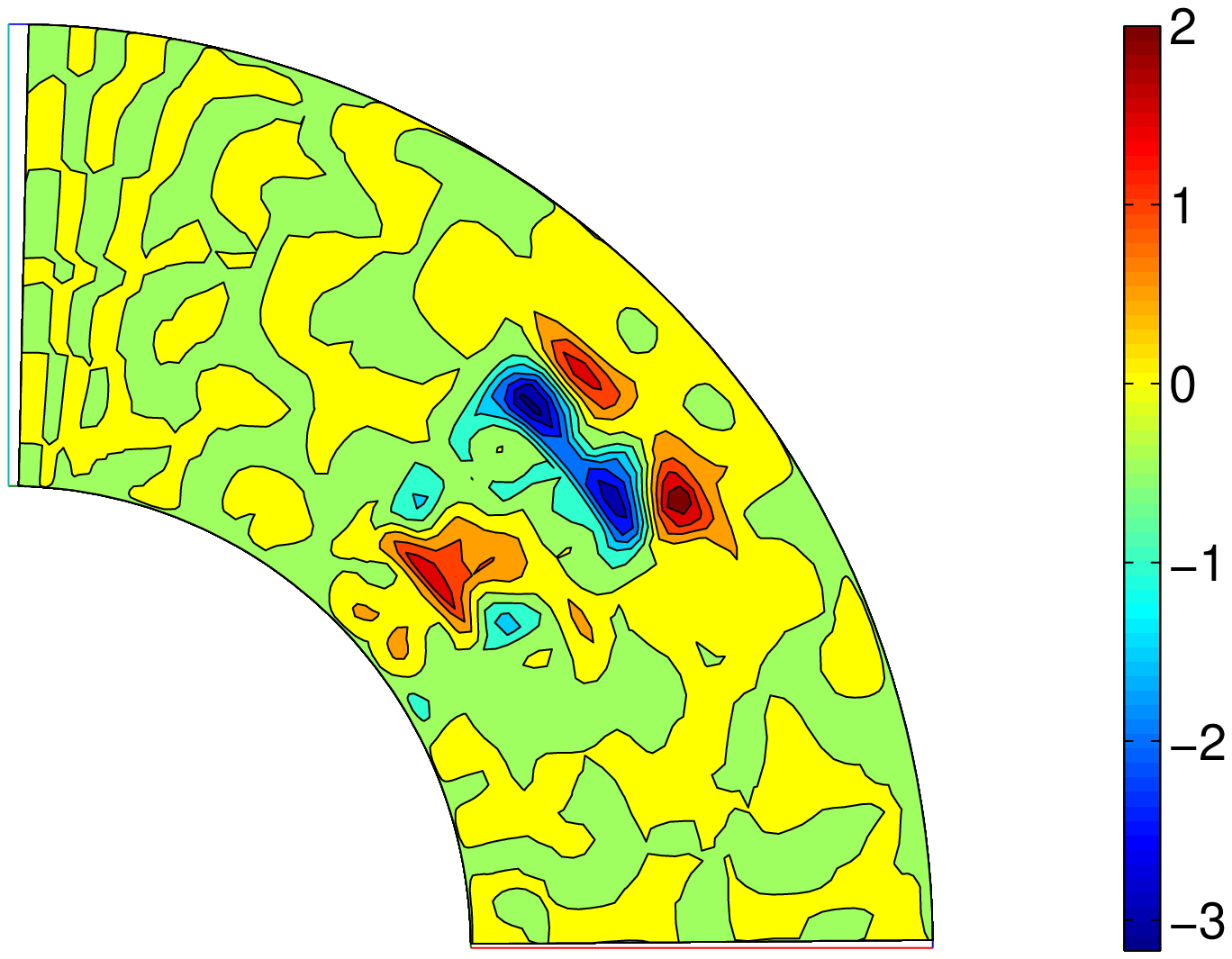}
}\\%
\subfigure[$N_r=10$, $S=0$]{%
\includegraphics[width=0.33\textwidth]{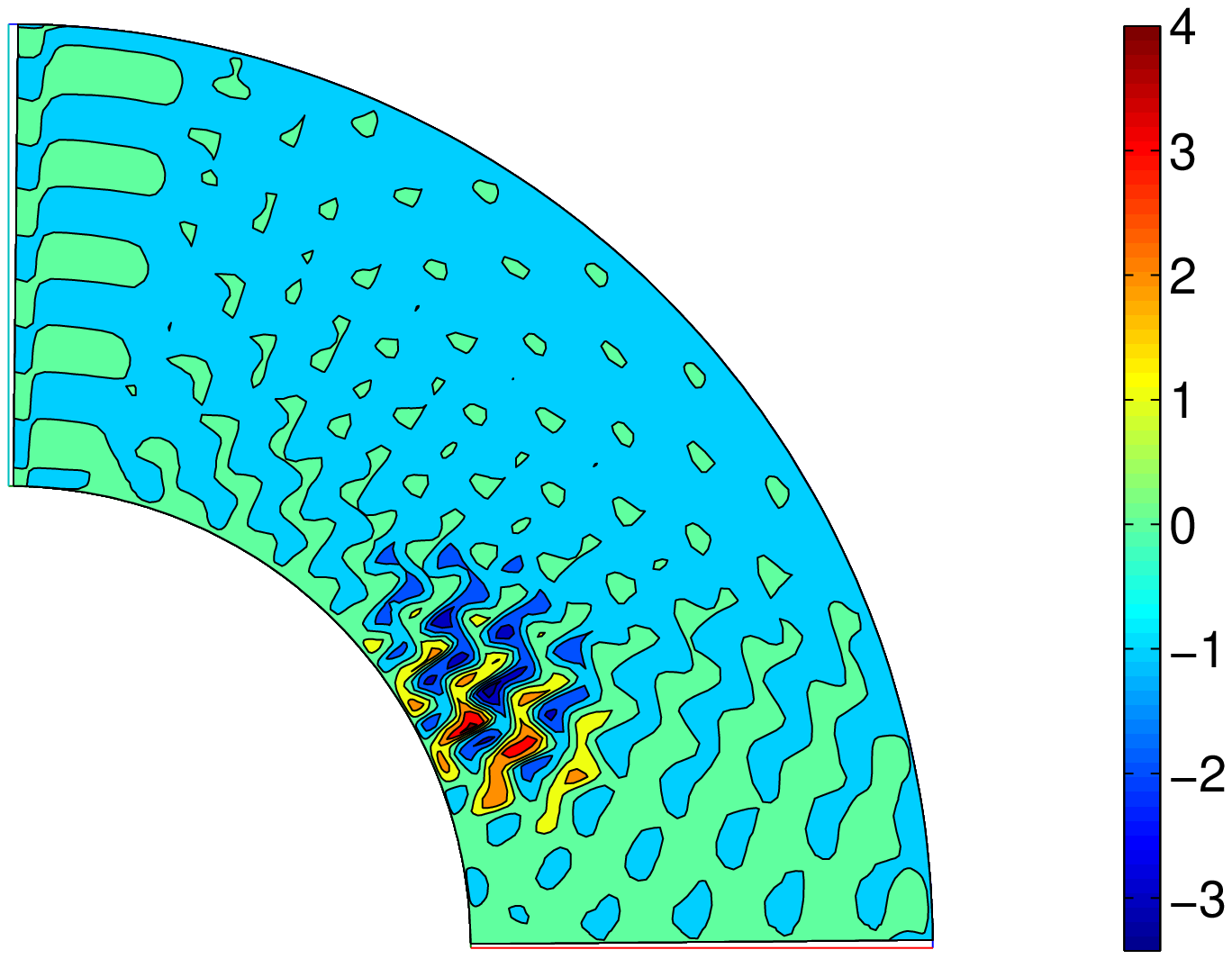}
}%
\subfigure[$N_r=10$, $S=3$, \textbf{$\Omega_1$}]{%
\includegraphics[width=0.34\textwidth]{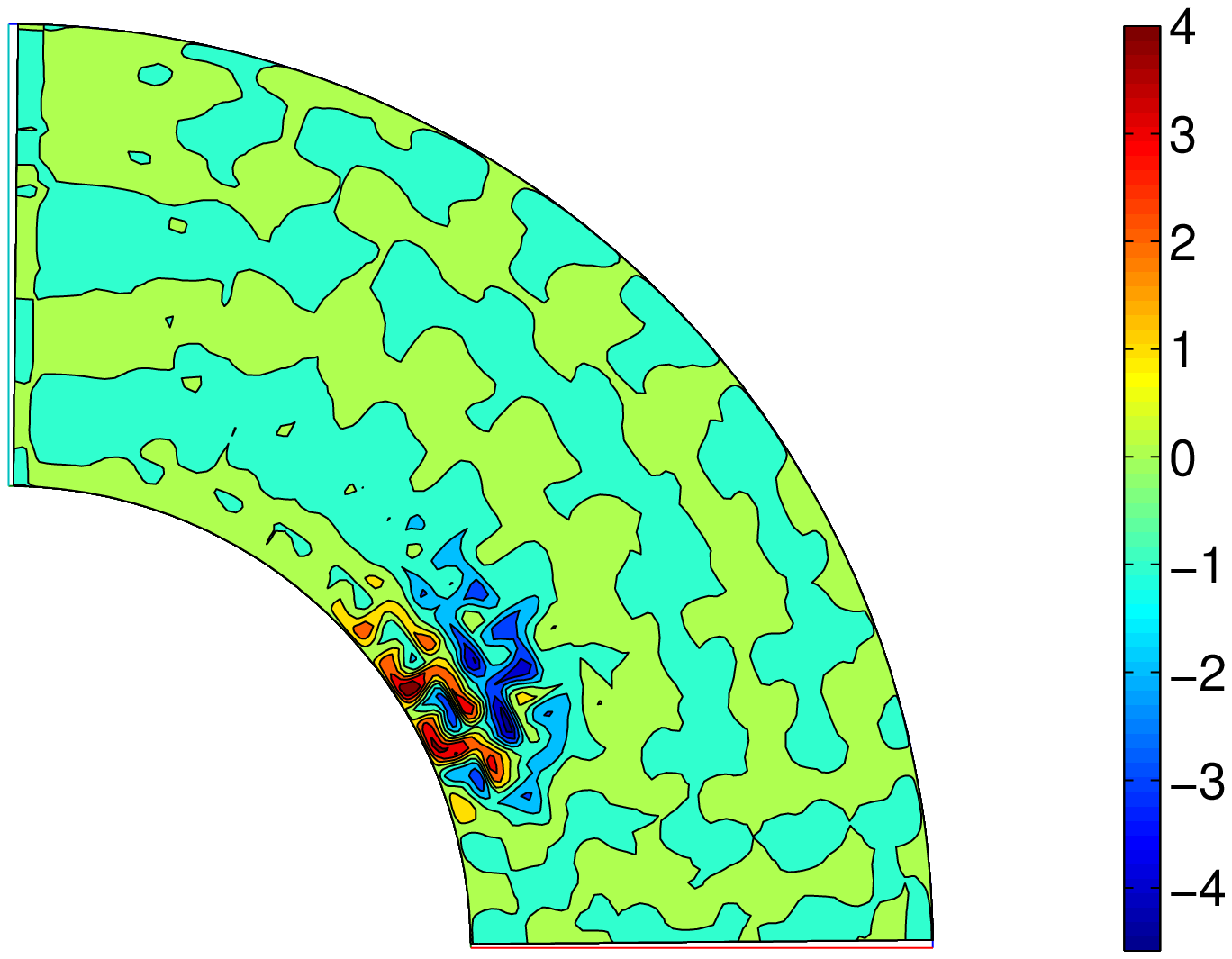}
}%
\subfigure[$N_r=10$, $S=3$, \textbf{$\Omega_2$}]{%
\includegraphics[width=0.34\textwidth]{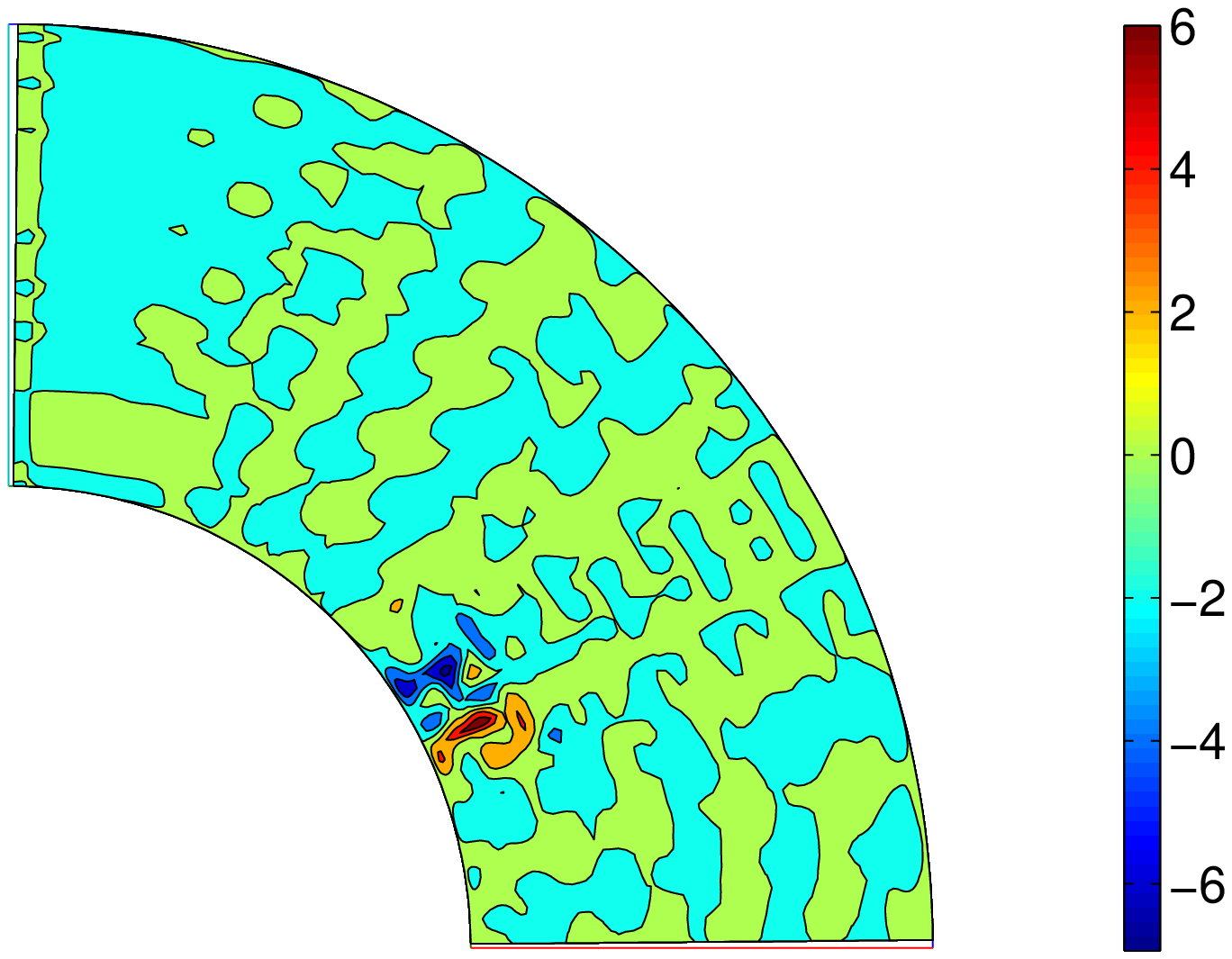}
}%
\caption{Meridional sections of $B_\phi$ for $N_r=5$ (top row) and $N_r=10$
(bottom row). From left to right $S=0$ (so either $\Omega_1$ or $\Omega_2$), then
$\Omega_1$ and $\Omega_2$, respectively, at $S=3$. $R_m =1000$ and $m=10$ in all cases.
The difference between $N_r=5$ and 10 is clearly visible in terms of the
different size structures, but the differences between $S=0$ and 3 are
surprisingly little, given how strongly suppressed the growth rates already
are.}
\label{fig:figure}
\end{figure}
Figure 3 shows examples of the spatial structure of the resulting eigenmodes.
As expected, the field is organized into structures on the scale of the
small-scale cells, and is also strongest at the mid-latitudes where the
helicity is strongest. Considering how strongly the growth rates vary, the
spatial structures vary surprisingly little.
\begin{figure}[ht!]
\centering
\subfigure[$N_r=5$, \textbf{$\Omega_1$}]{%
\includegraphics[width=0.49\textwidth]{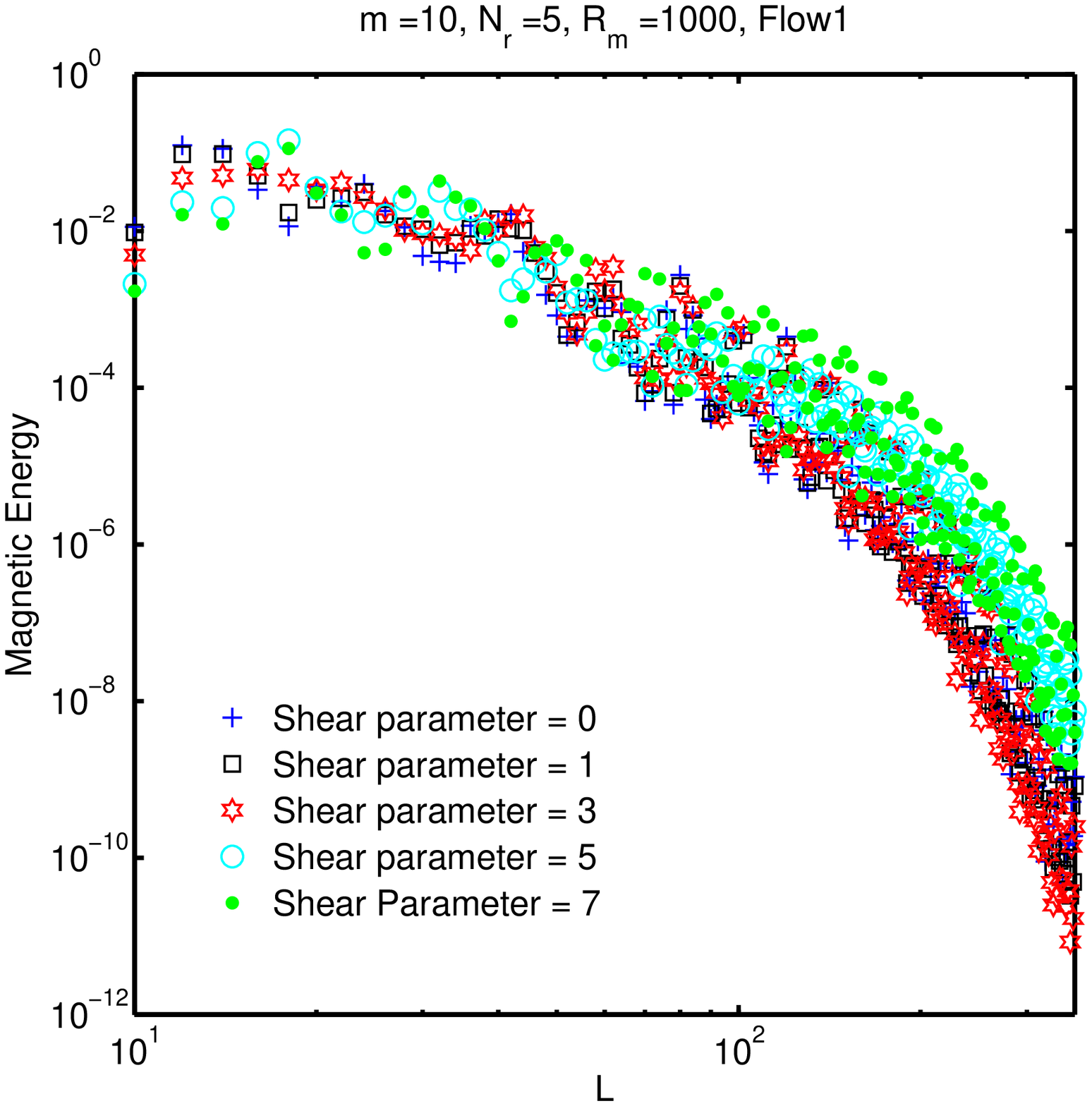}
}%
\subfigure[$N_r=5$, \textbf{$\Omega_2$}]{%
\includegraphics[width=0.5\textwidth]{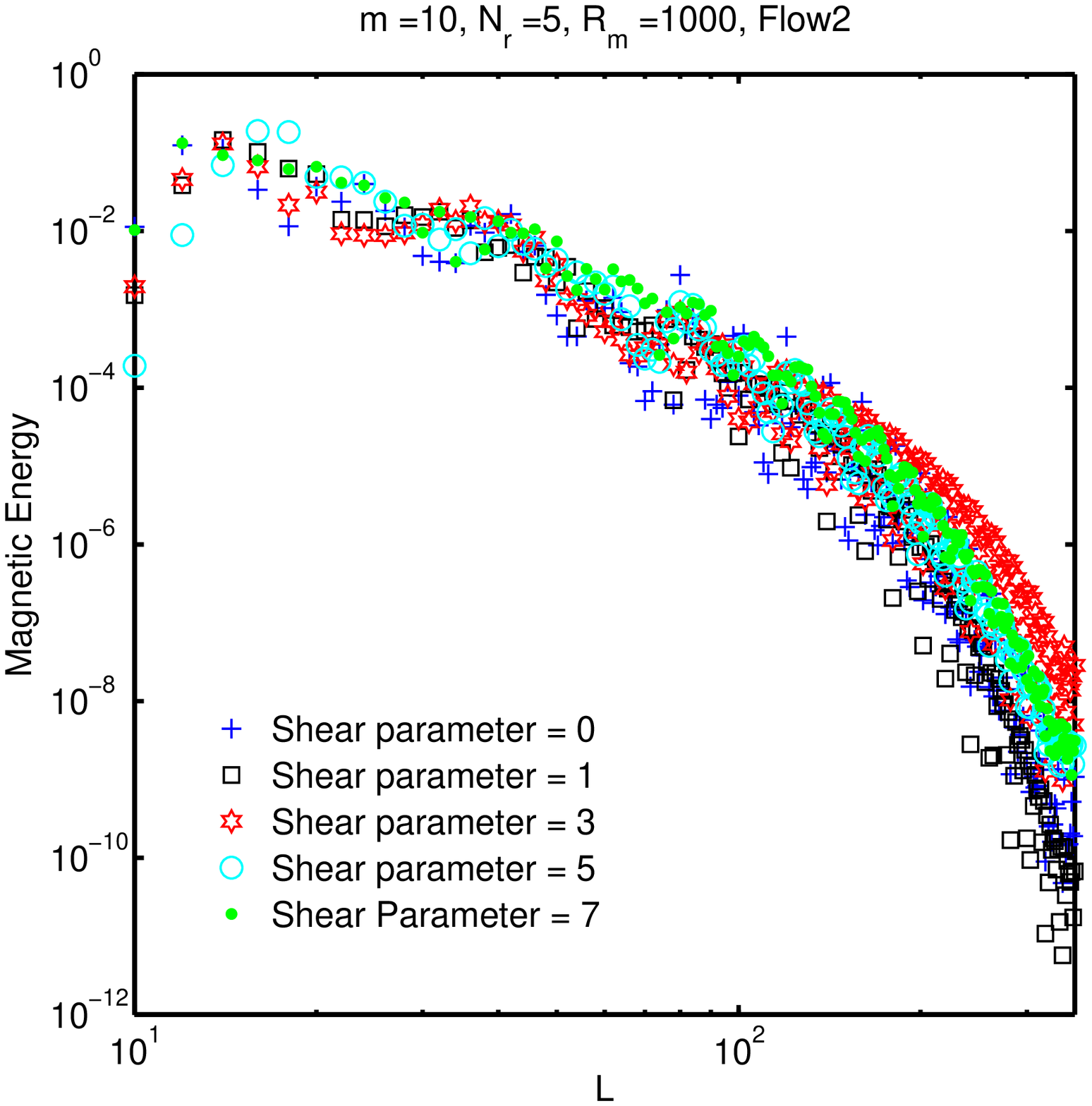}
}\\%
\subfigure[$N_r=10$, \textbf{$\Omega_1$}]{%
\includegraphics[width=0.5\textwidth]{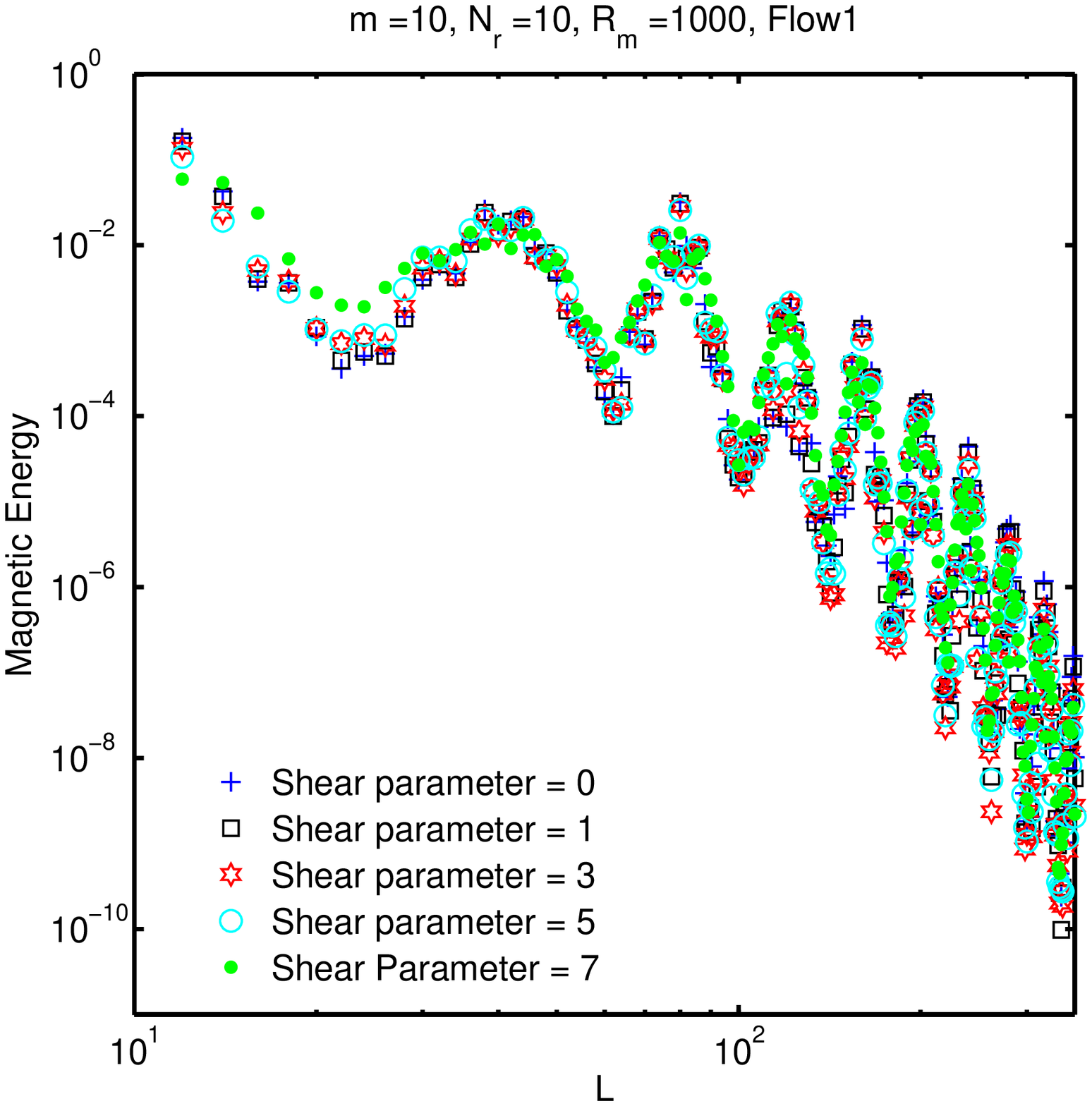}
}%
\subfigure[$N_r=10$, \textbf{$\Omega_2$}]{%
\includegraphics[width=0.5\textwidth]{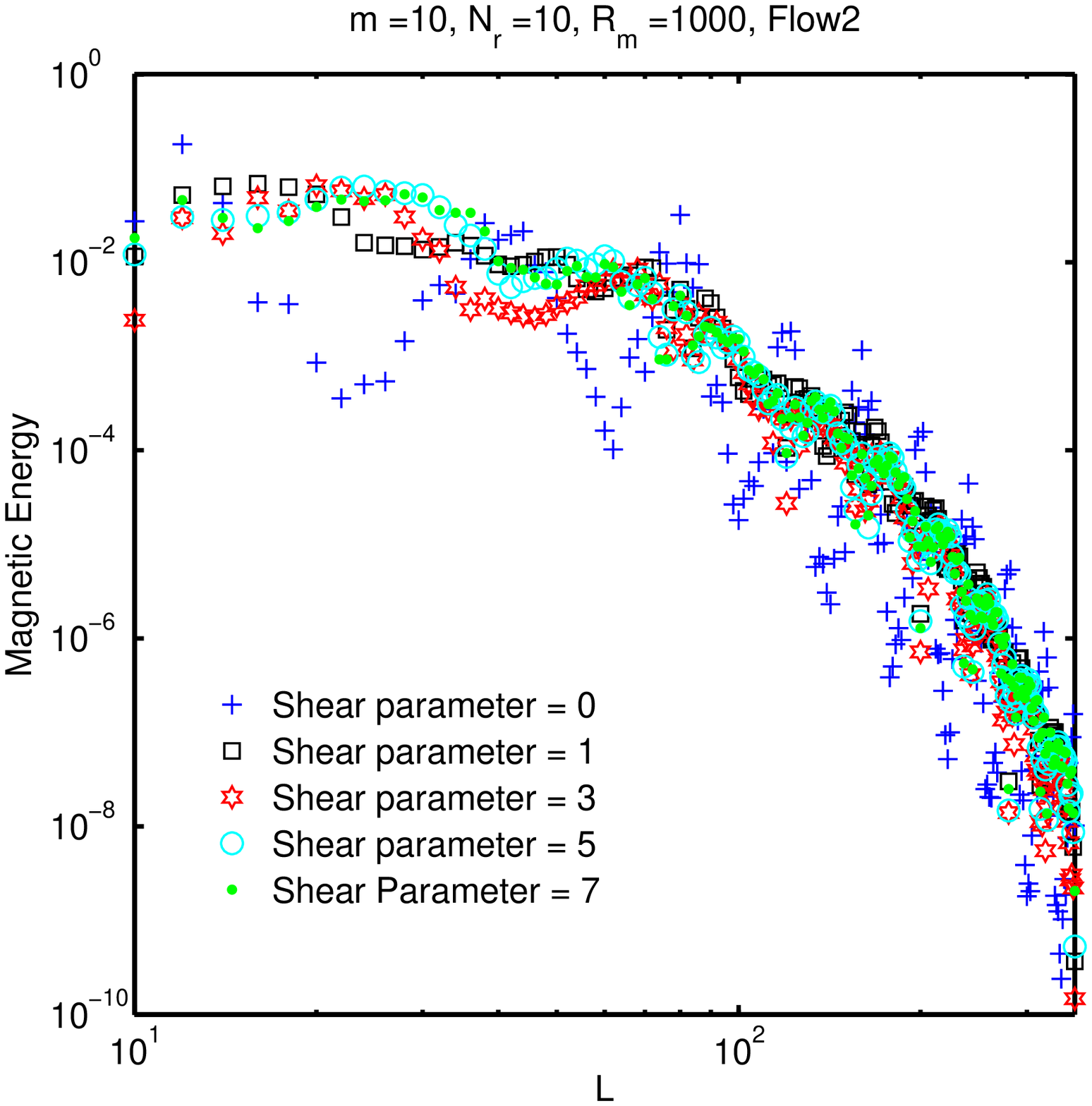}
}%
\caption{Energy spectra for magnetic field in the presence of shear $\Omega_1$ and $\Omega_2$ at $N_r=5$ and 10,
and $S=0,1,3,5,7$ as indicated.}
\end{figure}

Finally, Fig. 4 shows the energy spectra of these solutions in Fig. 3, as
well as other values of $S$.
Here we can see one clear difference between the two flows. Flow 1, with the
purely radial shear, exhibits strong peaks at spherical harmonics $l$ that
are multiples of $N_\theta$ (recall $N_\theta=20$ and 40 for $N_r=5$ and 10,
respectively), reflecting the number of small-scale cells. In contrast, for
Flow 2 with its latitudinal shear, these peaks have been largely smoothed out,
and one obtains much more uniform spectra. Another feature that is
particularly noticeable for $N_r=5$ is that increasing shear causes the
spectra to drop off less rapidly; that is, shear promotes small-scale in
the magnetic field, in agreement with \cite{KIM1,leprovost2009}.

\section{Conclusion}

We have seen in this work that at least for the small-scale flows considered
here, the addition of a large-scale shear always suppresses the dynamo
efficiency. There is thus no regime where the direct $\omega$-effect
dominates over the more indirect disrupting influences of the shear. One
important further direction for future work will be to take the small-scale
flows to be time-dependent, either periodic as in the \cite{Galloway}
plane-layer model, or stochastic as in a variety of analytic models
\cite{leprovost2008,leprovost2009,KIM3}, and see whether a large-scale shear
still always has a negative influence. Time-dependent flows will necessarily
involve additional correlation times and phase relationships, and may thus
yield different results for at least some parameter combinations.

Another possible extension would be to consider the nonlinear equilibration
of these dynamos, which would however require reintroducing the dynamics of
$\bf U$, and hence involve considerable complications over the problem
considered here. It is known (e.g.\ \cite{Phil}) that dynamically
equilibrated magnetic fields can be quite different from the kinematic
eigenfunctions, so potentially very rich additional dynamics could emerge.

\end{document}